\documentclass[12pt]{iopart}

\usepackage[T1]{fontenc}
\usepackage[latin1]{inputenc}
\usepackage{graphicx}
\usepackage{amssymb}

\makeatletter

\usepackage{epsfig}

\usepackage{dcolumn}

\usepackage{bm}

\usepackage{bbm}

\usepackage{wasysym}

\usepackage{marvosym}

\usepackage{hyperref}

\newlength{\textwidthm}

\makeatother

\begin{document}

\title{First order ferromagnetic phase transition in the low electronic density
regime of a biased graphene bilayer}

\author{T.~Stauber$^{1}$, Eduardo~V.~Castro$^{2}$, N.~A.~P.~Silva$^{1}$, and N.~M.~R.~Peres$^{1}$}

\address{$^{1}$Centro de F\'{\i}sica e Departamento de F\'{\i}sica, Universidade do Minho, P-4710-057, Braga, Portugal}

\address{$^{2}$ CFP and Departamento de F\'{\i}sica, Faculdade de Ci\^{e}ncias
Universidade do Porto, P-4169-007 Porto, Portugal}

\date{\today}

\begin{abstract}
The phase diagram of a biased graphene bilayer is computed and the
existence of a ferromagnetic phase is discussed both in the critical
on-site interaction $U_{c}$ versus doping density and versus
temperature.  We show that in the ferromagnetic phase the two planes
have unequal magnetization and that the electronic density is hole
like in one plane and electron like in the other. We give evidence for
a \emph{first-order} phase transition between paramagnetic and
ferromagnetic phases induced by doping at zero temperature.
\end{abstract}


\maketitle

\section{Introduction}

Graphene, a two-dimensional hexagonal lattice of carbon atoms, has
attracted considerable attention due to its unusual electronic properties,
characterized by massless Dirac Fermions.\cite{GN07,Kts06rev,NGPpw}
It was first produced via micromechanical cleavage on top of a SiO$_{2}$
substrate\cite{NGM+04,pnas} and its hallmark is the half integer
quantum Hall effect.\cite{NGM+05,ZTS+05}

\begin{figure}[htf]

\begin{centering}
\includegraphics[clip,width=7.5cm]{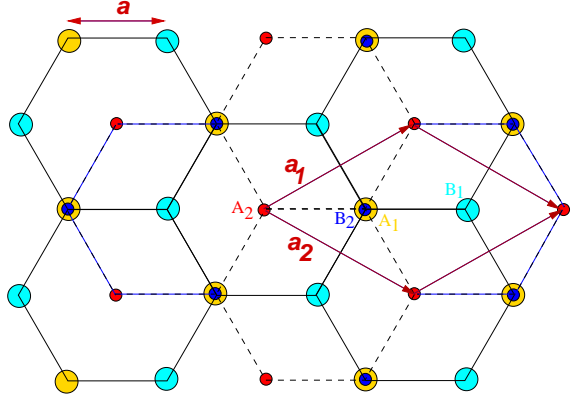} 
\par\end{centering}

\caption{(Color online) The unit cell a of graphene bilayer in the Bernal stacking.
The dashed hexagons are on top of the solid ones. The unit cell vectors
have coordinates $\bm a_{1}=a(3,\sqrt{3})/2$ and $\bm a_{2}=a(3,-\sqrt{3})/2.$}

\label{Fig_bilayer} 
\end{figure}

In addition to graphene, few-layer graphene can also be produced.
Of particular interest to us is the double layer graphene system,
where one encounters two carbon layers placed on top of each other
according to usual Bernal stacking of graphite (see Fig.~\ref{Fig_bilayer}).
The low-energy properties of this so-called bilayer graphene are then
described by massive Dirac Fermions.\cite{MF06} These new quasi-particles
have a quadratic dispersion close to the neutrality point and have
recently been identified in Quantum Hall measurements\cite{NMcCM+06}
and in Raman spectroscopy.\cite{FMS+06,GME+06}

In a graphene bilayer it is possible to have the two planes at different
electrostatic potentials.\cite{OBS+06,CNM+06} As a consequence,
a gap opens up at the Dirac point and the low energy band acquires
a Mexican hat relation dispersion.\cite{GNP06} This system is called
a biased graphene bilayer. The potential difference created between
the two layers can be obtained by applying a back gate voltage to
the bilayer system and covering the exposed surface with some chemical
dopant, as for example Potassium\cite{OBS+06} or NH$_{3}$.\cite{CNM+06}
In addition, it is also possible to control the potential difference
between the layers by using back and top gate setups.\cite{OHL+07}
The opening of the gap at the Dirac point in the biased bilayer system
was demonstrated both by angle resolved photoemission experiments
(ARPES)\cite{OBS+06} and Hall effect measurements.\cite{CNM+06}
The electronic gap in the biased system has been also observed in
epitaxially grown graphene films on SiC crystal surfaces.\cite{BSL+04}

Due to the Mexican hat dispersion relation the density of states close
to the gap diverges as the square root of the energy. The possibility
of having an arbitrary large density of states at the Fermi energy
poses the question whether this system can be unstable toward a
ferromagnetic ground state. The question of magnetism in carbon based
systems already has a long history. Even before the discovery of
graphene, highly oriented pyrolytic graphite (HOPG) has attracted a
broad interest due to the observation of anomalous properties, such as
magnetism and insulating behavior in the direction perpendicular to
the planes.  \cite{ESH+02,KEK02,KSE+03,KTS+03,OTH+07} The research of
$s-p$ based magnetism\cite{RGC+04,TNS+91,SDL+98} was especially
motivated by the technological use of nanosized particles of graphite,
which show interesting features depending on their shape, edges, and
applied field of pressure.\cite{EK05}

Microscopic theoretical models of bulk carbon magnetism include
nitrogen-carbon compositions where ferromagnetic ordering of spins
could exist in $\pi$ delocalized systems due to a lone electron pair
on a trivalent element\cite{Ovch78} or intermediate graphite-diamond
structures where the alternating $sp^{2}$ and $sp^{3}$ carbon atoms
play the role of different valence elements. \cite{OS91} More general
models focus on the interplay between disorder and
interaction. \cite{SGV05,VSS+05} Further, midgap states due to zig-zag
edges play a predominant role in the formation of magnetic moments
\cite{japonese,PCM+07} which support flat-band
ferromagnetism.\cite{Mielke91,Tasaki98,KM03} A generic model based on
midgap states was recently proposed in
Refs. \cite{WSG08,WSSG08}. Magnetism is also found in fullerene based
metal-free systems.\cite{CMG+04} For a recent overview on metal-free
Carbon-based magnetism see Ref.~\cite{MPbook06}.

To understand carbon-based magnetism in graphite, one may start with
the simplest case of one-layer, i.e., graphene. Because the density
of states of intrinsic graphene vanishes at the Dirac point, the simple
Stoner-like argument predicts an arbitrary large value of the Coulomb
on-site energy needed to produce a ferromagnetic ground state.\cite{PAB04,AP06}
In fact, because of the vanishing density of states, the Coulomb interaction
is not screened and the Hubbard model is not a good starting point
to study ferromagnetism in clean graphene. One, therefore, has to
consider the exchange instability of the Dirac gas due to the bare,
long-range Coulomb interaction in two dimensions. This study shows
that for a clean, doped or undoped graphene layer, a spin-polarized
ground-state due to the gain of exchange energy is only favorable
for unphysical values of the dimensionless coupling constant of graphene.\cite{PGN05}
The paramagnetic ground-state of clean graphene is thus stable against
the exchange interaction. If the system is disordered, e.g., due to
vacancies or edge states, a finite density of states builds in at
the Dirac point. As a consequence, a finite Hubbard interaction for
driving the system to a ferromagnetic ground state is obtained.\cite{PGN06}
In this case, the exchange interaction favors a ferromagnetic ground
state for reasonable values of the dimensionless coupling parameter.\cite{PGN05}
The presence of itinerant magnetism due to quasi-localized states
induced by single-atom defects in graphene, such as vacancies,\cite{PGS+06}
has also been obtained recently using first-principles.\cite{YH07}

The situation is quite different in a bilayer system. There, a finite
density of states exists at the neutrality point producing some amount
of screening in the system. Moreover, in the case of a biased bilayer
and for densities close to the energy gap, the density of states is
very large producing very effective screening. As a consequence, for
this system the Hubbard model is a good starting point to study the
tendency toward ferromagnetism. From the point of view of the exchange
instability of the bilayer system, it is found that the system is
always unstable toward a ferromagnetic ground state for low enough
particle densities. \cite{NNP+05,Sta07,CPSS08}

In the present paper, we want to explore the fact that the Hubbard
model is a good starting point to describe the Coulomb interactions
in the regime where the Fermi energy is close to the band edge of
the biased bilayer system. In particular we want to study the phase
diagram of the system as function of the doping. We further want to
determine the mean field critical temperature.

The paper is organized as followed. In section \ref{sec_model}, we
introduce the model and define the mean-field decoupling which allows
for different electronic density and magnetization in the two layers.
In section \ref{sec_ground_state}, we set up the mean-field equations
and present the numerical results in section \ref{sec_results}. We
close with conclusions and future research directions.


\section{Model Hamiltonian and mean field approximation}

\label{sec_model}

The Hamiltonian of a biased bilayer Hubbard model is the sum of two
pieces $H=H_{TB}+H_{U}$, where $H_{TB}$ is the tight-binding part
and $H_{U}$ is the Coulomb on-site interaction part of the Hamiltonian.
The tight-binding Hamiltonian is itself a sum of four terms describing
the tight-binding Hamiltonian of each plane, the hopping term between
the planes, and the electrostatic bias applied to the two planes.
We therefore have, \begin{equation}
H_{TB}=\sum_{\iota=1}^{2}H_{TB,\iota}+H_{\perp}+H_{V}\,,\end{equation}
 with \begin{eqnarray}
H_{TB,\iota}= & - & t\sum_{\bm R,\sigma}[a_{\iota\sigma}^{\dag}(\bm R)b_{\iota\sigma}(\bm R)+a_{\iota\sigma}^{\dag}(\bm R)b_{\iota\sigma}(\bm R-\bm a_{1})\nonumber \\
 & + & a_{\iota\sigma}^{\dag}(\bm R)b_{\iota\sigma}(\bm R-\bm a_{2})+H.c.]\,,\end{eqnarray}

\begin{equation}
H_{\perp}=-t_{\perp}\sum_{\bm R,\sigma}[a_{1\sigma}^{\dag}(\bm R)b_{2\sigma}(\bm R)+b_{2\sigma}^{\dag}(\bm R)a_{1\sigma}(\bm R)]\,,\end{equation}
 and \begin{equation}
H_{V}=\frac{V}{2}\sum_{\bm R,\sigma}[n_{a1\sigma}(\bm R)+n_{b1\sigma}(\bm R)-n_{a2\sigma}(\bm R)-n_{b2\sigma}(\bm R)]\,.\label{HV}\end{equation}
 As regards the bias term in Eq.~(\ref{HV}), we assume here that~$V$
can be externally controlled and is independent of the charge density
in the system. This situation can be realized with a back and top
gate setup.\cite{OHL+07} The on-site Coulomb part is given by, \begin{eqnarray}
H_{U}= & U & \sum_{\bm R}[n_{a1\uparrow}(\bm R)n_{a1\downarrow}(\bm R)+n_{b1\uparrow}(\bm R)n_{b1\downarrow}(\bm R)\nonumber \\
 & + & n_{a2\uparrow}(\bm R)n_{a2\downarrow}(\bm R)+n_{b2\uparrow}(\bm R)n_{b2\downarrow}(\bm R)]\,,\end{eqnarray}
 where $n_{x\iota\sigma}(\bm R)=x_{\iota\sigma}^{\dag}(\bm R)x_{\iota\sigma}(\bm R)$,
with $x=a,b$, $\iota=1,2$ and $\sigma=\uparrow,\downarrow$. 
%


The problem defined by the Hamiltonian $H_{TB}+H_{U}$ can not be
solved exactly and therefore we have to rest upon some approximation.
Here we adopt a mean field approach, neglecting quantum fluctuations.
Since we are interested in studying the existence of a ferromagnetic
phase we have to introduce a broken symmetry ground state. There is
however an important point to remark: since the two planes of the
bilayer are at different electrostatic potentials one should expect
that the electronic density and the magnetization will not be evenly
distributed among the two layers. Therefore our broken symmetry ground
state must take this aspect into account. As a consequence we propose
the following broken symmetry ground state: \begin{equation}
\langle n_{x1\sigma}(\bm R)\rangle=\frac{n+\Delta n}{8}+\sigma\frac{m+\Delta m}{8}\,,\end{equation}
 and \begin{equation}
\langle n_{x2\sigma}(\bm R)\rangle=\frac{n-\Delta n}{8}+\sigma\frac{m-\Delta m}{8}\,,\end{equation}
 where $n$ is the density per unit cell and $m=n_{\uparrow}-n_{\downarrow}$
is the spin polarization per unit cell. The quantities $\Delta n$
and $\Delta m$ represent the difference in the electronic density
and in the spin polarization between the two layers, respectively.\cite{foot1}
We note that $m$ and $\Delta m$ are independent parameters, being
in principle possible to have a ground state where $m=0$ but $\Delta m\ne0$.

\begin{figure}[htf]

\begin{centering}
\includegraphics[clip,width=7.5cm]{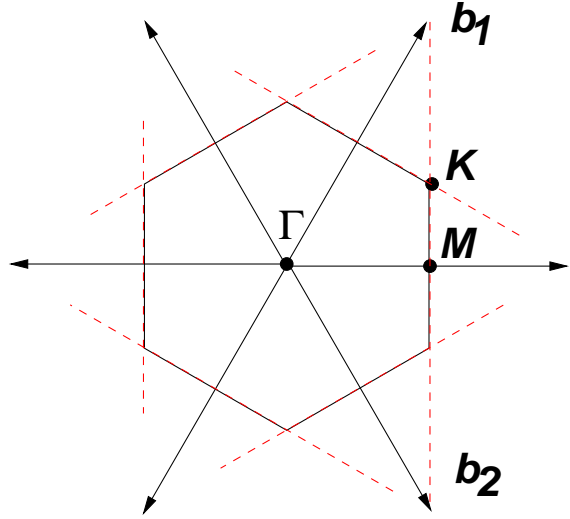} 
\par\end{centering}

\caption{(Color online) Brillouin zone of the bilayer problem. The Dirac point
$\bm K$ has coordinates $2\pi(1,\sqrt{3}/3)/(3a)$ and the $\bm M$
point has coordinates $2\pi(1,0)/(3a)$.}

\label{Fig_BZ} 
\end{figure}

When transformed to momentum space the mean field Hamiltonian obtained
from the above reads \begin{eqnarray}
H_{MF} & = & \sum_{\bm k,\sigma}\Psi_{\bm k,\sigma}^{\dag}H_{\bm k,\sigma}\Psi_{\bm k,\sigma}\nonumber \\
 & - & \frac{N_{c}U}{32}[(n+\Delta n)^{2}-(m+\Delta m)^{2}]\nonumber \\
 & - & \frac{N_{c}U}{32}[(n-\Delta n)^{2}-(m-\Delta m)^{2}]\,,\label{Eq_Hamilt}\end{eqnarray}
 with $\Psi_{\bm k,\sigma}^{\dag}=[a_{1\bm k\sigma}^{\dag},b_{1\bm k\sigma}^{\dag},a_{2\bm k\sigma}^{\dag},b_{2\bm k\sigma}^{\dag}]$
and $H_{\bm k,\sigma}$ given by \begin{equation}
H_{\bm k,\sigma}=\left(\begin{array}{cccc}
s_{\sigma} & -t\phi_{\bm k} & 0 & -t_{\perp}\\
-t\phi_{\bm k}^{\ast} & s_{\sigma} & 0 & 0\\
0 & 0 & p_{\sigma} & -t\phi_{\bm k}\\
-t_{\perp} & 0 & -t\phi_{\bm k}^{\ast} & p_{\sigma}\end{array}\right)\,,\end{equation}
 with $s_{\sigma}=\frac{V}{2}+\left(\frac{n+\Delta n}{8}-\sigma\frac{m+\Delta m}{8}\right)U$,
$p_{\sigma}=-\frac{V}{2}+\left(\frac{n-\Delta n}{8}-\sigma\frac{m-\Delta m}{8}\right)U$,
and $\phi_{\bm k}=1+e^{i\bm k\cdot\bm a_{1}}+e^{i\bm k\cdot\bm a_{2}}$.
The energy eigenvalues are given by, \begin{eqnarray}
E_{\sigma}^{j,l}(\bm k,m,\Delta m)&=&\left(\frac{n}{8}-\sigma\frac{m}{8}\right)U\\&+&
\frac{l}{2}\sqrt{2t_{\perp}^{2}+V_{\sigma}^{2}+4t^{2}\vert\phi_{\bm k}\vert^{2}+j2\sqrt{t_{\perp}^{4}+4t^{2}(t_{\perp}^{2}+V_{\sigma}^{2})\vert\phi_{\bm k}\vert^{2}}}\,,\nonumber\end{eqnarray}
 where $l,j=\pm$ and $V_{\sigma}$ is given by \begin{equation}
V_{\sigma}=V+U\Delta\tilde{n}-\sigma U\Delta\tilde{m}\,,\label{Vs}\end{equation}
 where we have introduced the definitions $\Delta n=4\Delta\tilde{n}$
and $\Delta m=4\Delta\tilde{m}$. It is clear that as long as $\Delta n$
and $\Delta m$ are finite the system has an effective $V_{\sigma}$
that differs from the bare value $V$. The momentum values are given
by, \begin{equation}
\bm k=\frac{m_{1}}{N}\bm b_{1}+\frac{m_{2}}{N}\bm b_{2}\,,\label{Eq_mom}\end{equation}
 with $m_{1},m_{2}=0,1,\ldots\, N-1$, the number of unit cells given
by $N_{c}=N^{2}$, and $\bm b_{1}$ and $\bm b_{2}$ given by, \begin{equation}
\bm b_{1}=\frac{2\pi}{3a}(1,\sqrt{3})\,,\hspace{0.5cm}\bm b_{2}=\frac{2\pi}{3a}(1,-\sqrt{3})\,.\end{equation}
 The Brillouin zone of the system is represented in Fig.~\ref{Fig_BZ}.
%



\section{Free energy and mean-field equations}

\label{sec_ground_state}

The free energy per unit cell, $f$, of Hamiltonian~(\ref{Eq_Hamilt})
is given by, \begin{eqnarray}
f=&-&\frac{k_{B}T}{N_{c}}\sum_{\bm k,\sigma}\sum_{l,j=\pm}\ln\left(1+e^{-(E_{\sigma}^{l,j}(\bm k)-\mu)/(k_{B}T)}\right)\\\nonumber
&-&\frac{U}{16}\big[n^{2}-m^{2}+(\Delta n)^{2}-(\Delta m)^{2}\big]+\mu n\,,\label{Eq_free_energy}\end{eqnarray}
 where $\mu$ is the chemical potential.

Let us introduce the density of states per spin per unit cell $\rho(E)$
defined as \begin{equation}
\rho(E)=\frac{1}{N_{c}}\sum_{\bm k}\delta(E-t\vert\phi_{\bm k}\vert)\,.\label{Eq_rho}\end{equation}
 The momentum integral in Eq. (\ref{Eq_rho}) is over the Brillouin
zone defined in Fig. \ref{Fig_BZ}, using the momentum definition
(\ref{Eq_mom}). The integral can be performed leading to, \begin{equation}
\rho(E)=\frac{2E}{t^{2}\pi^{2}}\left\{ \begin{array}{ccc}
\frac{1}{\sqrt{F(E/t)}}\mathbf{K}\left(\frac{4E/t}{F(E/t)}\right)\,, &  & 0<E<t\,,\\
\\\frac{1}{\sqrt{4E/t}}\mathbf{K}\left(\frac{F(E/t)}{4E/t}\right)\,, &  & t<E<3t\,,\end{array}\right.\label{eq:DOS1L}\end{equation}
 where $F(x)$ is given by \begin{equation}
F(x)=(1+x)^{2}-\frac{(x^{2}-1)^{2}}{4}\,,\label{eq:Fe}\end{equation}
 and $\mathbf{K}(m)$ is defined as, \begin{equation}
\mathbf{K}(m)=\int_{0}^{1}dx[(1-x^{2})(1-mx^{2})]^{-1/2}\,.\label{eq:K}\end{equation}

Using Eq.~(\ref{Eq_rho}), the free energy in Eq.~(\ref{Eq_free_energy})
can be written as a one-dimensional integral, \begin{eqnarray}
f & =\nonumber \\
- & k_{B}T & \sum_{\sigma}\sum_{l,j=\pm}\int dE\rho(E)\ln\left(1+e^{-(E_{\sigma}^{l,j}(E)-\mu)/(k_{B}T)}\right)\nonumber \\
 & - & \frac{U}{16}\big[n^{2}-m^{2}+(\Delta n)^{2}-(\Delta m)^{2}\big]+\mu n\,.\label{Eq_FE}\end{eqnarray}
 The mean field equations are now obtained from the minimization of
the free energy (\ref{Eq_FE}). The doping, $\delta n$, relative
to the situation where the system is at half filling is defined as,
\begin{eqnarray}
\delta n=\sum_{\sigma}\sum_{l,j=\pm}\int dE\rho(E)f[E_{\sigma}^{l,j}(E)-\mu]-4\,,\label{Eq_MF1}\end{eqnarray}
 where $f(x)=(1+e^{x/(k_{B}T)})^{-1}$. The spin polarization per
unit cell obeys the mean field equation, \begin{eqnarray}
m=\sum_{\sigma}\sum_{l,j=\pm}\sigma\int dE\rho(E)f[E_{\sigma}^{l,j}(E)-\mu]\,.\label{Eq_MF2}\end{eqnarray}
 The difference in the electronic density between the two layers is
obtained from, \begin{equation}
\Delta\tilde{n}=\frac{1}{2}\sum_{\sigma}\sum_{l,j=\pm1}\int dE\rho(E)f[E_{\sigma}^{l,j}(E)-\mu]v_{\sigma}^{l,j}(E)\,,\label{eq:Dn}\end{equation}
 where $v_{\sigma}^{l,j}(E)$ is given by \begin{equation}
v_{\sigma}^{l,j}(E)=\frac{l}{2}\frac{V_{\sigma}}{\sqrt{\ldots}}\left(1+\frac{j4E^{2}}{\sqrt{t_{\perp}^{4}+4E^{2}(t_{\perp}^{2}+V_{\sigma}^{2})}}\right)\,,\end{equation}
 and \begin{equation}
\sqrt{\ldots}=\sqrt{2t_{\perp}^{2}+V_{\sigma}^{2}+4E^{2}+j2\sqrt{t_{\perp}^{4}+4E^{2}(t_{\perp}^{2}+V_{\sigma}^{2})}}\,.\end{equation}
 The difference in the magnetization between the two layers is obtained
from \begin{equation}
\Delta\tilde{m}=\frac{1}{2}\sum_{\sigma}\sum_{l,j=\pm1}\sigma\int dE\rho(E)f[E_{\sigma}^{l,j}(E)-\mu]v_{\sigma}^{l,j}(E)\,,\label{eq:Dm}\end{equation}

Let us now assume that the system supports a ferromagnetic ground
state whose magnetization vanishes at some critical value $U_{c}$
at zero temperature. Additionally we assume that $\Delta m=0$ when
$m=0$, which will be shown to be the case in this system. The value
of $U_{c}$ is determined from expanding (\ref{Eq_MF2}) to first
order in $m$, leading to, \begin{eqnarray}
1 & = & \frac{U_{c}}{4}\sum_{l,j=\pm1}\int dE\rho(E)\delta[E_{\sigma}^{l,j}(E,0,0)-\mu]\nonumber \\
 & = & \frac{U_{c}}{4}\sum_{l,j,k=\pm1}^{\ast}\frac{\rho(E_{k}^{\ast})}{\vert f'_{l,j}(E_{k}^{\ast})\vert}\theta(3t-E_{k}^{\ast})\theta(E_{k}^{\ast})\nonumber \\
 & = & \frac{U_{c}}{4}\rho_{b}(\tilde{\mu},U_{c})=U_{c}\tilde{\rho}_{b}(\tilde{\mu},U_{c})\,,\label{Eq_stoner}\end{eqnarray}
 where $\rho_{b}(\tilde{\mu},U_{c})$ is the density of states per
unit cell per spin 
for a biased bilayer at the energy $\tilde{\mu}=\mu-nU_{c}/8$ and
$\tilde{\rho}_{b}(\tilde{\mu},U_{c})$ is the density of states per
spin per lattice point. Although Eq.~(\ref{Eq_stoner}) looks like
the usual Stoner criterion the fact that 
the bias $V_{\sigma}$ given in Eq.~(\ref{Vs}) depend on $U$ due
to the difference in the electronic density $\Delta n$ makes Eq.~(\ref{Eq_stoner})
a non-linear equation for $U_{c}$ which must be solved self-consistently.
For low doping $\delta n$ the product $U_{c}\Delta\tilde{n}$ is
a small number when compared to $V$ and therefore it can be neglected
in Eq.~(\ref{Vs}). In this case Eq.~(\ref{Eq_stoner}) reduces
to the usual Stoner criterion: \begin{equation}
U_{c}\simeq1/[\tilde{\rho}_{b}(\tilde{\mu})]\,.\label{scapp}\end{equation}

The quantities $E_{k}^{\ast}$ in Eq.~(\ref{Eq_stoner}) are the
roots of the delta function argument, \begin{equation}
E_{\sigma}^{l,j}(E_{k}^{\ast})-\mu=0\,.\label{eq:roots_eq}\end{equation}
 The quantity $f'_{l,j}(E_{k}^{\ast})$ is the derivative in order
to the energy $E$ of Eq.~(\ref{eq:roots_eq}) evaluated at the roots
$E_{k}^{\ast}$. The roots $E_{k}^{\ast}$ are given by \begin{equation}
E_{k}^{\ast}=\frac{1}{2}\sqrt{4\tilde{\mu}^{2}+V_{\sigma}^{2}+k2\sqrt{4\tilde{\mu}^{2}(t_{\perp}^{2}+V_{\sigma}^{2})-t_{\perp}^{2}V_{\sigma}^{2}}}\,,\label{roots}\end{equation}
 with $k=\pm$. Equation~(\ref{eq:roots_eq}) cannot be solved for
all bands: the existence of a solution is determined by $\mu$. As
a consequence we added the~$\ast$ symbol in the summation of Eq.~(\ref{Eq_stoner}),
which means that only bands for which Eq.~(\ref{eq:roots_eq}) can
be solved (two at the most) contribute to the summation. It also means
that for the contributing bands only real roots in Eq.~(\ref{roots})
are taken into account to the summation. The number of real roots
in Eq.~(\ref{roots}) depends on the particular band an $\mu$ through
Eq.~(\ref{eq:roots_eq}). As the function $f'_{l,j}(E)$ is given
by \begin{equation}
f'_{l,j}(E)=\frac{2lE}{\sqrt{\ldots}}\left(1+j\frac{t_{\perp}^{2}+V_{\sigma}^{2}}{\sqrt{t_{\perp}^{4}+4E^{2}(t_{\perp}^{2}+V_{\sigma}^{2})}}\right)\,.\end{equation}
 it is clear that both roots are imaginary for $\tilde{\mu}$ in the
range \begin{equation}
-\frac{t_{\perp}V_{\sigma}}{2\sqrt{t_{\perp}^{2}+V_{\sigma}^{2}}}<\tilde{\mu}<\frac{t_{\perp}V_{\sigma}}{2\sqrt{t_{\perp}^{2}+V_{\sigma}^{2}}}\,,\end{equation}
 which means that the system has an energy gap of value \begin{equation}
\Delta_{g}=\frac{t_{\perp}V_{\sigma}}{\sqrt{t_{\perp}^{2}+V_{\sigma}^{2}}}\,.\label{gap}\end{equation}
 We finally note that since we have assumed $\Delta m=0$, $V_{\sigma}$
does not effectively depend on $\sigma$.


\section{Results and Discussion}

\label{sec_results}

We start with the zero temperature phase diagram in the plane $U$~vs.~$\delta n$.
An approximate analytic treatment is possible in this limit, which
is used to check our numerical results. The effect of temperature
is considered afterwards.


\subsection{Zero temperature}

\label{subsec_zeroT}


\subsubsection{Approximate solution}

\label{subsubsec_appT0} 
\begin{figure}[t]

\begin{centering}
\includegraphics[clip,width=0.9\columnwidth]{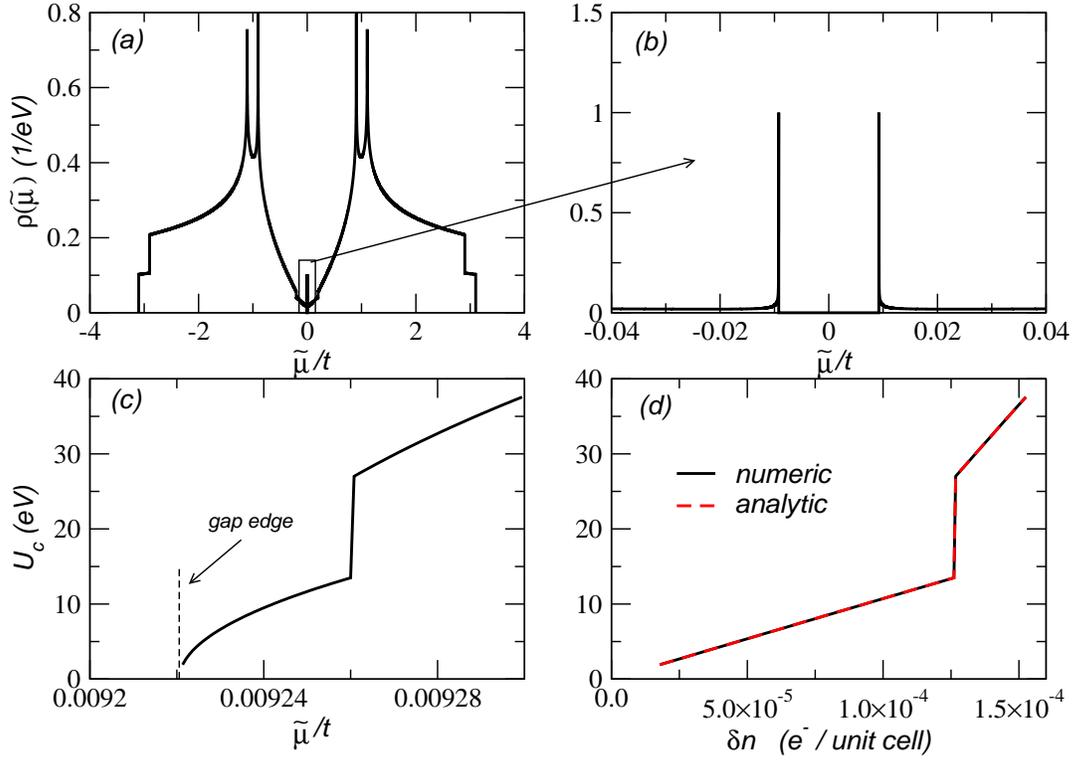} 
\par\end{centering}

\caption{(Color online) (a)~--~Density of states $\rho(\tilde{\mu})$ per
unit cell per spin of the bilayer problem with $U=0$. (b)~--~Zoom
of~(a) near the gap region. (c)~--~Critical value $U_{c}$ for
ferromagnetism in the low doping, $\delta n$, regime. (d)~--~The
same as in~(c) as a function of doping. The parameters are $t=2.7$~\texttt{eV},
$t_{\perp}=0.2t$, $V=0.05$~\texttt{eV}. The edge of the gap is
located at $\Delta_{g}/(2t)\simeq0.00922$. }

\label{Fig_DOS} 
\end{figure}

In Fig.~\ref{Fig_DOS} we represent the
density of states of a biased bilayer with $U=0$ together with the
low doping critical value $U_{c}$, as given by Eq.~(\ref{scapp}).
In panel~(b) of Fig.~\ref{Fig_DOS} a zoom in of the density of
states close to the gap is shown. It is clear that the density of
states diverges at the edge of the gap. As consequence the closer
to edge of the gap the chemical potential is the lower will be the
critical $U_{c}$ value. This quantity is shown in panel~(c) of Fig.~\ref{Fig_DOS}
as function of the chemical potential $\tilde{\mu}$ and in panel~(d)
as a function of doping $\delta n$. The lowest represented value
of $U_{c}$ is about $U_{c}\simeq2.7$ \texttt{eV} to which corresponds
an electronic doping density $\delta n\simeq2.5\times10^{-5}$ electrons
per unit cell. The step like discontinuity shown in panels~(c) and~(d)
for $U_{c}$ occurs when the Fermi energy equals $V/2$, signaling
the top of the Mexican hat dispersion relation.

%


It is clear from panel~(d) of Fig.~\ref{Fig_DOS} that in the low
doping limit $U_{c}$ is a linear function of doping $\delta n$.
This limit enables for an approximate analytic treatment which not
only explains the linear behavior but also provides a validation test
of our numerical results. Firstly we note that for very low doping
the density of states in Eq.~(\ref{scapp}) is close to the gap edge,
$|\tilde{\mu}|\sim\Delta_{g}/2$, where $\Delta_{g}$ is the size
of the gap Eq.~(\ref{gap}). In this energy region the density of
states has a 1D like divergence,\cite{GNP06} behaving as, \begin{equation}
\rho_{b}(\tilde{\mu})\propto\frac{1}{\sqrt{|\tilde{\mu}|-\Delta_{g}/2}}\,.\label{dos1D}\end{equation}
 Using this approximate expression to compute the doping, $\delta n\propto\textrm{sign}(\tilde{\mu})\times\int_{\Delta_{g}/2}^{|\tilde{\mu}|}\textrm{d}x~\rho_{b}(x)$,
we immediately get $\delta n\propto\textrm{sign}(\tilde{\mu})/\rho_{b}(\tilde{\mu})$
and thus $U_{c}\propto|\delta n|$. In order to have an analytic expression
for $U_{c}$ in the low doping limit we have to take into account
the proportionality coefficient in Eq.~(\ref{dos1D}). After some
algebra it can be shown that the density of states per spin per lattice
point near the gap edge can be written as, \begin{equation}
\rho_{b}(\tilde{\mu})\approx\frac{1}{t^{2}4\pi^{2}}\sqrt{\frac{\Delta_{g}(t_{\perp}^{2}+V^{2})}{F(\chi)}}\mathbf{K}\left(\frac{4\chi}{F(\chi)}\right)\frac{1}{\sqrt{|\tilde{\mu}|-\Delta_{g}/2}}\,,\label{eq:dosapp}\end{equation}
 where $\chi=[(\Delta_{g}^{2}+V^{2})/(4t^{2})]^{1/2}$, with $F(x)$
and $\mathbf{K}(m)$ as in Eqs.~(\ref{eq:Fe}) and~(\ref{eq:K}).
The doping $\delta n$, measured with respect to half filling in units
of electrons per unit cell, can be written as, \begin{eqnarray}
\delta n & =&\textrm{sign}(\tilde{\mu})\times8\int_{\Delta_{g}/2}^{|\tilde{\mu}|}\mbox{d}x\,\rho_{b}(x)\nonumber \\
 & \approx&\frac{4}{t^{2}\pi^{2}}\sqrt{\frac{\Delta_{g}(t_{\perp}^{2}+V^{2})}{F(\chi)}}\mathbf{K}\left(\frac{4\chi}{F(\chi)}\right)\sqrt{|\tilde{\mu}|-\Delta_{g}/2}.\label{eq:dnapp}\end{eqnarray}
 Inserting Eq.~(\ref{eq:dosapp}) into Eq.~(\ref{scapp}), and taking
into account Eq.~(\ref{eq:dnapp}), we are able to write, \begin{equation}
U_{c}\approx t^{4}\pi^{4}\frac{F(\chi)}{\Delta_{g}(t_{\perp}^{2}+V^{2})}\left[\mathbf{K}\Big(\frac{4\chi}{F(\chi)}\Big)\right]^{-2}\delta n\,.\label{eq:Ucapp}\end{equation}
 In panel~(d) of Fig.~\ref{Fig_DOS} both the numerical result of
Eq.~(\ref{scapp}) and the analytical result of Eq.~(\ref{eq:Ucapp})
are shown. The agreement is excellent.

%
\begin{figure}[t]

\begin{centering}
\includegraphics[clip,width=0.98\columnwidth]{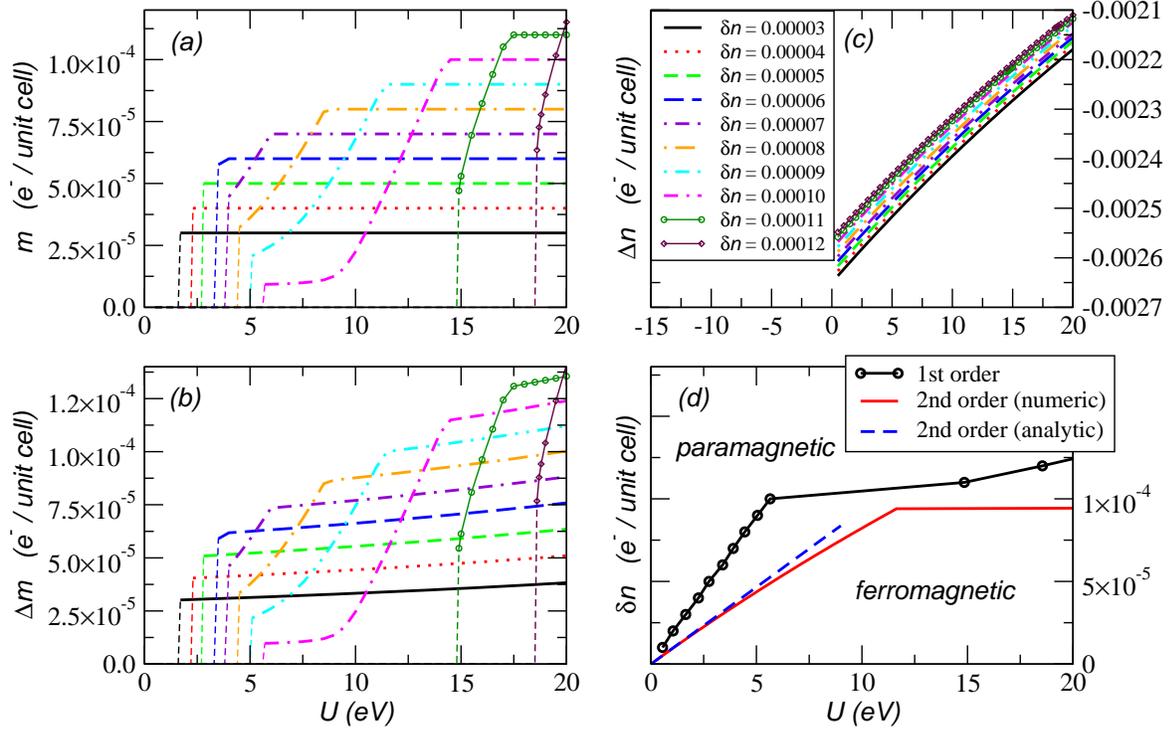} 
\par\end{centering}

\caption{(Color online) 
Panels~(a), (b), and~(c) show the zero temperature self-consistent
solution for~$m$, $\Delta m$, and $\Delta n$, respectively. The
zero temperature phase diagram of the biased bilayer in the $U$ vs.
$\delta n$ plane is shown in panel~(d). Symbols in panel~(d) are
inferred from panel~(a) and signal a \textit{first-order} phase transition;
the solid {[}Eq.~(\ref{Eq_stoner})] and dashed {[}Eq.~(\ref{eq:Ucapp})]
lines stand for a \textit{second-order} phase transition. The constant
parameters are $V=0.05$ \texttt{eV}, $t_{\perp}=0.2t$, and $t=2.7$
\texttt{eV}.}

\label{Fig_PD_n} 
\end{figure}



\subsubsection{Self-consistent solution}

\label{subsubsec_scT0} We now need to solve the mean field equations
in order to obtain the zero temperature phase diagram of the biased
bilayer. 
In order to achieve this goal we study how $m$, $\Delta m$, and
$\Delta n$ depend on the interaction $U$, for given values of the
electronic doping $\delta n$.

In panel~(a) of Fig.~\ref{Fig_PD_n} it is shown how $m$ depends
on $U$ for different values of $\delta n$. The chosen values of
$\delta n$ correspond to the chemical potential being located at
the divergence of the low energy density of states. The lower the
$\delta n$ is the more close to the gap edge is the chemical potential
and therefore the larger the density of states is. As a consequence,
$m$ presents a smaller critical $U_{c}$ value for smaller $\delta n$
values. It is interesting to note that the magnetization saturation
values correspond to full polarization of the doping charge density
with $m=\delta n$, also found within a one-band model.\cite{Sta07}
In panel~(b) of Fig.~\ref{Fig_PD_n} we plot the $\Delta m$ mean
field parameter. Interestingly the value of $\Delta m$ vanishes at
the same $U_{c}$ as $m$. For finite values of $m$ we have $\Delta m>m$,
which means that the magnetization of the two layers is opposite.
We therefore have two ferromagnetic planes that possess opposite and
unequal magnetization. In panel~(c) of Fig.~\ref{Fig_PD_n} we show
the value of $\Delta n$ as function of $U$. It is clear that $|\delta n|<|\Delta n|$,
which implies that the density of charge carriers is above the Dirac
point in one plane and below it in the other plane. This means that
the charge carriers are electron like in one plane and hole like in
the other.

In panel~(d) of Fig.~\ref{Fig_PD_n} we show the phase diagram of
the system in the $U$ vs. $\delta n$ plane. Symbols are inferred
from the magnetization behavior in panel~(a). They signal a \textit{first-order}
phase transition when $m$ increases from zero to a finite value {[}see
panel~(a)]. The full (red) line is the numerical self-consistent
result of Eq.~(\ref{Eq_stoner}), and the dashed (blue) line is the
approximate analytic result given by Eq.~(\ref{eq:Ucapp}). The discrepancy
between lines and symbols has a clear meaning. In order to obtain
both Eqs.~(\ref{Eq_stoner}) and~(\ref{scapp}) we assumed that
a \textit{second-order} phase transition would take place, i.e., the
magnetization $m$ would vanish continuously when some critical $U_{c}$
is approached from above. This is not the case, and the system undergoes
a first-order phase transition for smaller $U$ values than those
for the second-order phase transition case. There are clearly two
different regimes in panel~(d) of Fig.~\ref{Fig_PD_n}: one at densities
lower than $\delta n\lesssim1\times10^{-4}$, where the dependence
of $\delta n$ on $U_{c}$ is linear, and another regime for $\delta n>1\times10^{-4}$
where a plateau like behavior develops. This plateau has the same
physical origin as the step like discontinuity we have seen in panels~(c)
and~(d) of Fig.~\ref{Fig_DOS}. Clearly, as the density $\delta n$
grows the needed value of $U_{c}$ for having a ferromagnetic ground
state increases. This is a consequence of the diverging density of
states close to the gap edge. As regards the limit $\delta n\rightarrow0$
it is obvious from panel~(d) of Fig.~\ref{Fig_PD_n} that we have
$U_{c}\rightarrow0$. It should be noted, however, that lowering the
density $\delta n$ leads to a decrease of $m$ and $\Delta m$, as
can be seen in panels~(a) and~(b) of Fig.~\ref{Fig_PD_n}. Therefore,
even though we have $U_{c}\rightarrow0$ in the limit $\delta n\rightarrow0$,
we have also $m\rightarrow0$ and $\Delta m\rightarrow0$, which implies
a \emph{paramagnetic} ground state for the undoped ($\delta n=0$)
biased bilayer. Only $\Delta n$ remains finite at zero doping, in
agreement with the observations that screening is still possible at
the neutrality point ($\delta n=0$ ).\cite{McC06,CNM+06,MSB+06}

\begin{figure}
\begin{centering}
\includegraphics[clip,width=0.98\columnwidth]{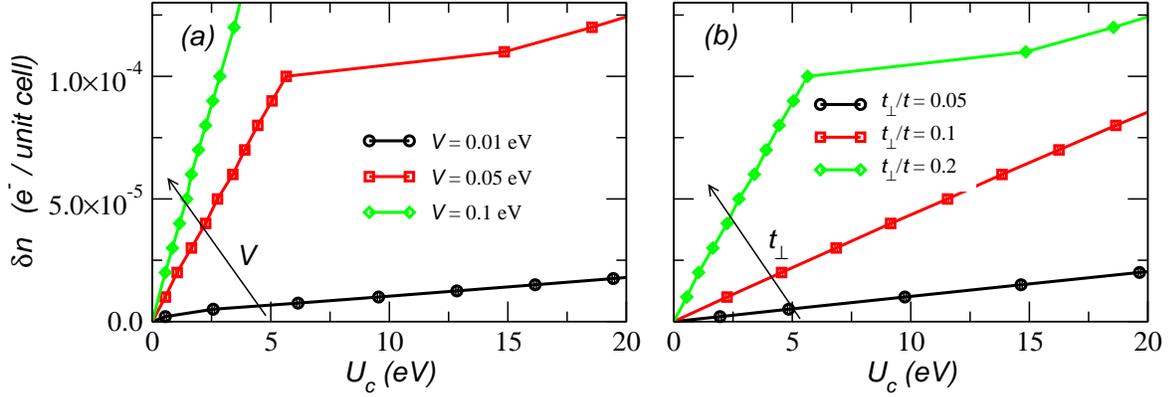} 
\par\end{centering}

\caption{\label{fig:PDgen} (Color online) Effect of $t_{\perp}$ and $V$
on the zero temperature $U_{c}$ vs. $\delta n$ phase diagram: (a)~fixed
$t_{\perp}=0.2t$ and varying $V$; (b)~fixed $V=0.05\,\mathtt{eV}$
and varying $t_{\perp}$. For a given $\delta n$ the \emph{ferromagnetic}
phase establishes for $U>U_{c}$ and the \emph{paramagnetic} phase
for $U<U_{c}$.}

\end{figure}

So far we have analyzed the system for fixed values of the bias voltage,
$V$, and interlayer coupling, $t_{\perp}$. In Fig.~\ref{fig:PDgen}
we show the effect of the variation of these two parameters on the
zero temperature phase diagram. In panel~(a) we have fixed the interlayer
coupling, $t_{\perp}=0.2t$, and varied the bias voltage, $V(\mathtt{eV})=\{0.01,0.05,0.1\}$;
in panel~(b) we did the opposite, with $V=0.05\,\mathtt{eV}$ and
$t_{\perp}/t=\{0.05,0.1,0.2\}$. Essentially, raising either $V$
or $t_{\perp}$ leads to a decrease of the critical interaction, $U_{c}$,
needed to establish the ferromagnetic phase for a given $\delta n$.
The order of the transition, however, remains \emph{first-order}:
for a given $\delta n$, the critical interaction $U_{c}$ predicted
by Eq.~(\ref{Eq_stoner}), which is valid for a \emph{second-order}
phase transition, is always higher than what is obtained by solving
self-consistently the mean-field equations, meaning that a \emph{first-order}
transitions is occurring at a lower $U_{c}$. It is interesting to
note that the effect of $V$ and $t_{\perp}$ on the \emph{first-order}
critical $U_{c}$ line is similar to what is expected for the usual
Stoner criterion, where increasing either $V$ or $t_{\perp}$ gives
rise to an increase in the density of states at the Fermi level and
a lower $U_{c}$ thereof.

The bias voltage and the interlayer coupling have also interesting
effects on the magnetization, $m$, and spin polarization difference
between layers, $\Delta m$. Decreasing $t_{\perp}$ leads to a decrease
in $\Delta m$, and below some $t_{\perp}$ we have $\Delta m<m$,
as opposed to the case discussed above ($V=0.05\,\mathtt{eV}$ and
$t_{\perp}=0.2t$). In particular, for $V=0.05\,\mathtt{eV}$, we
have already found $\Delta m<m$ for $t_{\perp}\leq0.1t$. A similar
effect has been observed when $V$ is increased. For $t_{\perp}=0.2t$
we have found $\Delta m<m$ for $V\geq0.1\,\mathtt{eV}$. It should
be noted, however, that $m$ and $\Delta m$ are $U$-dependent. Increasing
$U$ leads $m$ to saturate while $\Delta m$ keeps growing, as can
be seen in panels~(a) and~(b) of Fig.~\ref{Fig_PD_n} for the particular
case of $V=0.05\,\mathtt{eV}$ and $t_{\perp}=0.2t$. This means that,
depending on the value of the parameters $V$ and $t_{\perp}$, we
can go from $\Delta m<m$ to $\Delta m>m$ just by increasing the
interaction strength $U$. It can also be seen in panel~(a) of Fig.~\ref{Fig_PD_n}
that $m$ is completely saturated at the transition for $\delta n<\delta n_{c}\approx6\times10^{-5}$~electrons
per unit cell, while for $\delta n>\delta n_{c}$ it saturates only
at some $U>U_{c}$. Even though this behavior seems to be general
for any $V$ and $t_{\perp}$, the value of $\delta n_{c}$ is not.
In particular, we have found $\delta n_{c}$ to depend strongly on
$V$~--~it seems to vary monotonically with $V$, increasing when
$V$ increases. Let us finally comment on the effect of $V$ and $t_{\perp}$
on the charge imbalance between planes, $\Delta n$. Irrespective
of $V$ and $t_{\perp}$ we have always observed $|\delta n|<|\Delta n|$,
which means that charge carriers are always electron like in one plane
and hole like in the other. As expected, increasing/decreasing either
$V$ or $t_{\perp}$ leads to an increase/decrease of $\Delta n$.


\subsubsection{Understanding the asymmetry between planes}

\label{subsubsec_assymT0}

%
\begin{figure}[t]

\begin{centering}
\includegraphics[width=0.8\columnwidth]{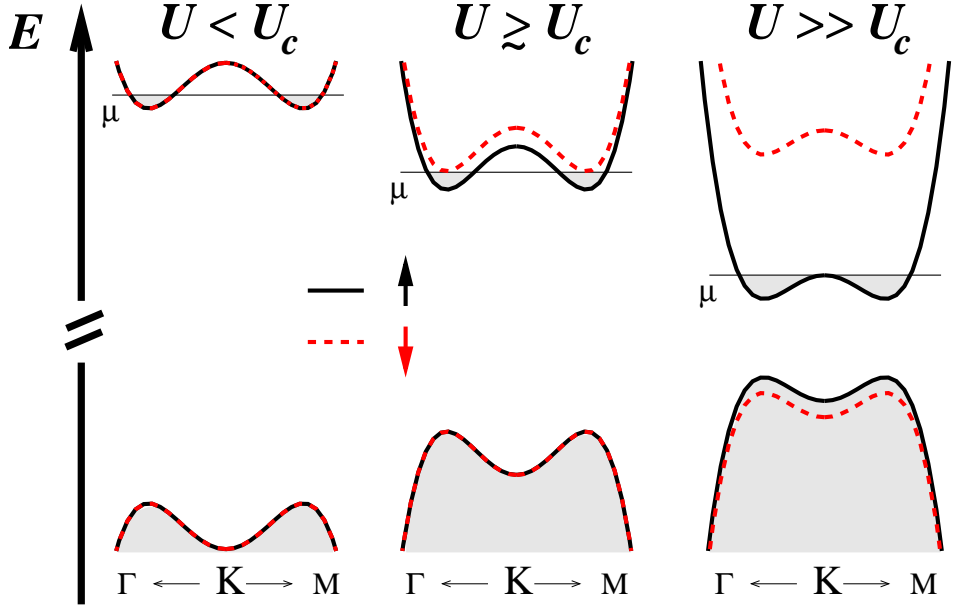} 
\par\end{centering}

\caption{\label{fig:bs} (Color online) Hartree-Fock bands for \emph{up} (full
lines) and \emph{down} (dashed lines) spin polarizations. Three different
cases are considered (from left to right): $U<U_{c}$, $U\gtrsim U_{c}$,
and $U\gg U_{c}$.}

\end{figure}

The asymmetry between planes regarding both charge and spin polarization
densities can be understood based on the Hartree-Fock bands shown
in Fig.~\ref{fig:bs}. The figure stands for $V=0.05\,\mathtt{eV}$
and $t_{\perp}=0.2t$, but can easily be generalized for other parameter
values.

It should be noted firstly that in the biased bilayer the weight of
the wave functions in each layer for near-gap states is strongly dependent
on their valence band or conduction band character.\cite{CNM+06,McC06,MSB+06}
Valence band states near the gap have their amplitude mostly localized
on layer~2, due to the lower electrostatic potential $-V/2$ {[}see
Eq.~(\ref{HV})]. On the other hand, near-gap conduction band states
have their highest amplitude on layer~1, due to the higher electrostatic
potential $+V/2$ for this layer {[}see Eq.~(\ref{HV})].

The case $U<U_{c}$ shown in Fig.~\ref{fig:bs} (left) stands for
the paramagnetic phase. The values $m=0$ and $\Delta m=0$ seen in
this phase are an immediate consequence of the degeneracy of \emph{up}
and \emph{down} spin polarized bands. The presence of a finite gap,
however, leads to the abovementioned asymmetry between near-gap valence
and conduction states. As a consequence, a half-filled bilayer would
have $n_{2}=(4+\Delta n)/2$ electrons per unit cell on layer~2 (electron
like charge carriers) and $n_{1}=(4-\Delta n)/2$ electrons per unit
cell on layer~1 (hole like charge carriers), with $\Delta n\neq0$.
Even though the system studied here is not at half-filling, as long
as $|\delta n|<|\Delta n|$ the carriers on layers~1 and~2 will
still be hole and electron like, respectively. Note that as $U$ is
increased the charge imbalance $\Delta n$ is suppressed in order
to reduce the system Coulomb energy, as can be seen in panel~(c)
of Fig.~\ref{Fig_PD_n}. From the band structure point of view a
smaller $\Delta n$ is the result of a smaller gap $\Delta_{g}$,
which means that increasing $U$ has the effect of lowering the gap.

Let us now consider the case $U\gtrsim U_{c}$ shown in Fig.~\ref{fig:bs}
(center). The degeneracy lifting of spin polarized bands gives rise
to a finite magnetization, $m\neq0$. Interestingly enough, the degeneracy
lifting is only appreciable for conduction bands, as long as $U$
is not much higher than $U_{c}$. This explains why the total polarization~$m$
and the difference in polarization between layers~$\Delta m$ have
similar values, $m\approx\Delta m$, as shown in panels~(a) and~(b)
of Fig.~\ref{Fig_PD_n}~--~as only conduction bands are contributing
to $\Delta m$, the spin polarization density is almost completely
localized in layer~1, where $m_{1}=(m+\Delta m)/2\approx m$, while
the spin polarization in layer~2 is negligible, $m_{2}=(m-\Delta m)/2\approx0$.

It is only when $U\gg U_{c}$ that valence bands become non-degenerate,
as seen in Fig.~\ref{fig:bs} (right). This implies that near-gap
valence states with \emph{up} and \emph{down} spin polarization have
different amplitudes in layer~2. As the valence band for \emph{down}
spin polarization has a lower energy the near-gap valence states with
spin \emph{down} have higher amplitude in layer~2 than their spin
\emph{up} counterparts. Consequently, the magnetization in layer~2
is effectively opposite to that in layer~1, i.e., $\Delta m>m$.
This can be observed in panels~(a) and~(b) of Fig.~\ref{Fig_PD_n},
where as $U$ is increased the magnetization of the two layers becomes
opposite.

We note, however, that the cases $U\gtrsim U_{c}$ and $U\gg U_{c}$
mentioned above are parameter dependent. For instance, the valence
bands can show an appreciable degeneracy lifting already for $U\gtrsim U_{c}$,
especially for small values of the $t_{\perp}$ parameter ($t_{\perp}\lesssim0.05t$).
In this case the magnetization of the two layers is no longer opposite,
with $\Delta m<m$. This can be understood as due to the fact that
as $t_{\perp}$ is decreased the weight of near-gap wave functions
becomes more evenly distributed between layers, leading not only to
a decrease in $\Delta n$ but also in $\Delta m$. As $U$ is further
increased the energy splitting between \emph{up} and \emph{down} spin
polarized bands gets larger, enhancing $\Delta m$. For $U\gg U_{c}$,
and depending on the parameters $V$ and $t_{\perp}$, the magnetization
of the two layers may become opposite even for small $t_{\perp}$
values.

%
\begin{figure}[t]

\begin{centering}
\includegraphics[width=0.98\columnwidth]{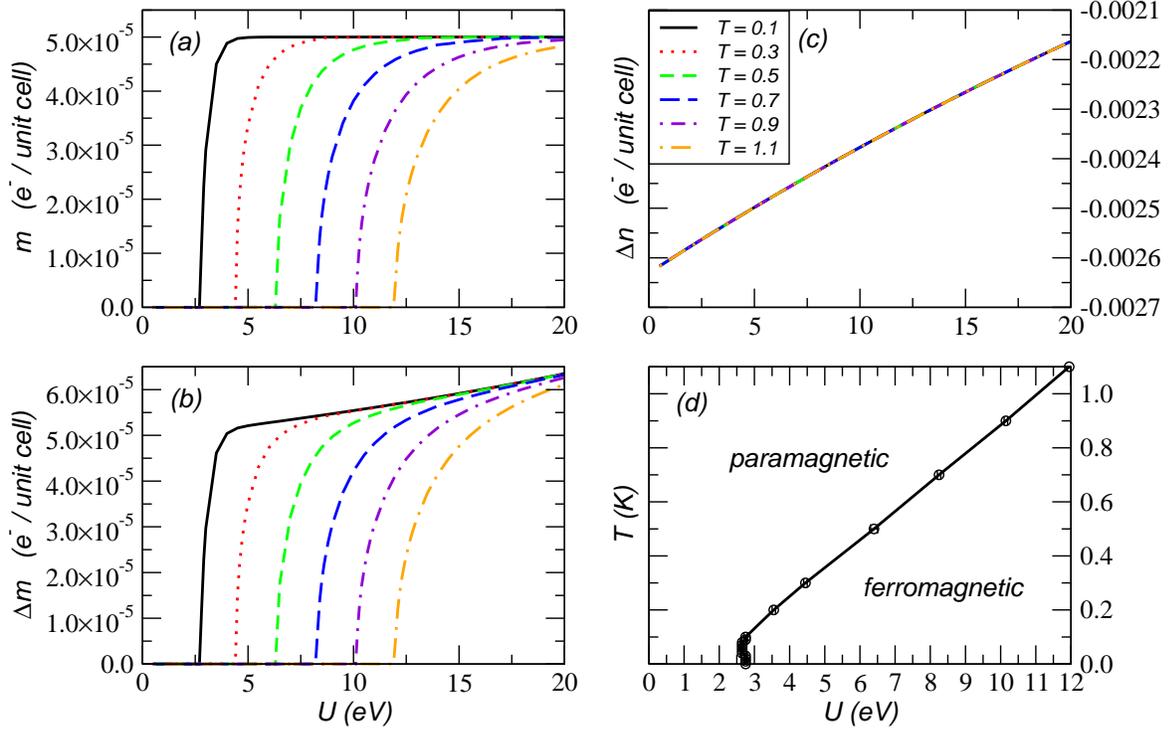} 
\par\end{centering}

\caption{(Color online) Panels~(a), (b), and~(c) show the finite temperature
self-consistent solution for~$m$, $\Delta m$, and $\Delta n$,
respectively, with temperature measured in~\texttt{K}. The finite
temperature phase diagram of the biased bilayer in the $U$ vs. $T$
plane is shown in panel~(d). The constant parameters are $V=0.05$
\texttt{eV}, $t_{\perp}=0.2t$, $t=2.7$ \texttt{eV}, and $\delta n=0.00005$
\texttt{e$^{-}$/unit cell}.}

\label{Fig_PD_T} 
\end{figure}


\subsection{Finite temperature}

\label{subsec_T}

Next we want to describe the phase diagram of the bilayer in the temperature
vs. on-site Coulomb interaction $U$ plane. This is done in Fig.~\ref{Fig_PD_T}
for a charge density $\delta n=5\times10^{-5}$ electrons per unit
cell. For temperatures ranging from zero to $T=$1.1~\texttt{K} we
studied the dependence of $m$, $\Delta m$ and $\Delta n$ on the
Coulomb on-site interaction $U$. First we note that the minimum critical
value $U_{c}$ is not realized at zero temperature. There is a reentrant
behavior which is signaled by the smallest $U_{c}$ for $T=0.06\pm0.02$\texttt{~K}.
For temperatures above $T\approx0.1$\texttt{~K} we have larger $U_{c}$
values for the larger temperatures, as can be seen in panel~(a).
The same is true for $\Delta m$, panel~(b). As in the case of Fig.~\ref{Fig_PD_n},
the value of $\Delta m$, at a given temperature and $U$ value, is
larger than $m$. Also the value of $\Delta n$, shown in panel~(c),
is larger than $\delta n$. Therefore we have the two planes presenting
opposite magnetization and the charge carriers being hole like in
one graphene plane and electron like in the other plane. In panel~(d)
of Fig.~\ref{Fig_PD_T} we present the phase diagram in the $T$
vs. $U$. Except at very low temperatures, there is a linear dependence
of $T$ on $U_{c}$. It is clear that at low temperatures, $T\simeq$
0.2\texttt{~K}, the value of $U_{c}$ is smaller than the estimated
values of $U$ for carbon compounds.\cite{Parr50,Baeriswyl86}

\subsection{Disorder}

Crucial prerequisite in order to find ferromagnetism is a high DOS at
the Fermi energy. The presence of disorder will certainly cause a
smoothing of the singularity in the DOS and the band gap
renormalization, and can even lead to the closing of the gap. We note,
however, that for small values of the disorder strength the DOS still
shows an enhanced behavior at the band gap edges.\cite{NN06,NNG+07}
The strong suppression of electrical noise in bilayer
graphene\cite{LA08} further suggests that in addition to a high
crystal quality -- leading to remarkably high mobilities\cite{MNK+07}
-- an effective screening of random potentials is at work. Disorder
should thus not be a limiting factor in the predicted low density
ferromagnetic state, as long as standard high quality BLG samples are
concerned.

Let us also comment on the next-nearest interlayer-coupling $\gamma_{3}$,
which in the unbiased case breaks the spectrum into four pockets for
low densities.\cite{MF06} In the biased case, $\gamma_{3}$ still
breaks the cylindrical symmetry, leading to the trigonal distortion
of the bands, but the divergence in the density of states at the edges
of the band gap is preserved.\cite{NNG+07} Therefore, the addition
of $\gamma_{3}$ to the model does not qualitatively change our result.

\section{Summary}

We have investigated the tendency of a biased bilayer graphene towards
a ferromagnetic ground state. For this, we used a mean-field theory
which allowed for a different carrier density and magnetization in
the two layers. We have found that in the ferromagnetic phase the
two layers have unequal magnetization and that the electronic density
is hole like in one plane and electron like in the other. We have
also found that at zero temperature, where the transition can be driven
by doping, the phase transition between paramagnetic and ferromagnetic
phases is \emph{first-order}.

\ack
TS, EVC, and NMRP acknowledge the financial support
from POCI 2010 via project PTDC/FIS/64404/2006, and the financial
support of Funda\c{c}\~ao para a Ci\^{e}ncia e a Tecnologia through Grant No.~SFRH/BD/13182/2003. Support from ESF via INSTANS is also acknowledged.

\section{References}

\bibliographystyle{unsrt}
\bibliography{grapheneexp,graphenetheo,myrefs,graphenerev,carbonmag,footnotes}

\begin{thebibliography}{10}
\bibitem{foot1}
Assuming equal spin densities in sublattices $A$ and $B$ of the same layer is a
  reasonable approximation for $t_{\perp}\ll t$.
\end{thebibliography}

\end{document}